\documentclass[epj]{svjour}

\usepackage{graphicx}
\usepackage{fancyhdr}
\def\makeheadbox{\flushright{\sc{HIP-2007-55/TH}}{%  remove EPJC header
 }}
\def\mytitle{My title} 
\def\myauthors{My name}  
\def\mytype{My type of session}
\def\mysession{My session}
\newcommand{\ba}{\begin{array}}
\newcommand{\ea}{\end{array}}
\newcommand{\bd}{\begin{displaymath}}
\newcommand{\ed}{\end{displaymath}}
\newcommand{\be}{\begin{equation}}
\newcommand{\ee}{\end{equation}}
\newcommand{\bea}{\begin{eqnarray}}
\newcommand{\eea}{\end{eqnarray}}

\def\q2 {q^2}

\def\miss {\hspace{-0.5cm}\slash~~}

\def\rslep {\tilde{e_R}}
\def\rsnu {\tilde{\nu}_R}
\def\snu {\tilde{\nu}}
\def\lslep {\tilde{e_L}}
\def\stau {\tilde{\tau}}
\def\mer {m_{\rslep}}
\def\mmr {m_{\tilde{\mu}_R}}
\def\mml {m_{\tilde{\mu}_L}}
\def\mel {m_{\lslep}}

\def\mytitle{Right-Chiral Sneutrino LSP in mSUGRA} %Put your title here!
\def\myauthors{Santosh Kumar Rai}    %Put your name here!
\def\mytype{Contributed Talk}    
\def\mysession{Colliders - SUSY Phenomenology}
\begin{document}
\title{Right-Chiral Sneutrino LSP in mSUGRA: Event characteristics of 
       NLSP at the LHC}
\author{Sudhir Kumar Gupta\inst{1}
  \and
  Biswarup Mukhopadhyaya\inst{1}
  \and 
  \underline{Santosh Kumar Rai}\inst{2}
  \fnmsep
  \thanks{\emph{Email:} santosh.rai@helsinki.fi}%
}
\institute{ Harish-Chandra Research Institute,
  Chhatnag Road, Jhusi,               
  Allahabad - 211 019, India
  \and
  Helsinki Institute of Physics
  and High Energy Physics Division,
  Department of Physical
  Sciences, PL 64, {FIN-00014},
  University of Helsinki, Finland
}
\date{}
\abstract{
We study a supersymmetric scenario where the lighter tau-sneutrino is the 
lightest supersymmetric particle, while the lighter stau-state is the next 
lightest. Such a scenario can be motivated within the framework of minimal 
supergravity, with just the addition of a right-chiral neutrino superfield. 
Such a spectrum leads to rather unusual signals of supersymmetry, showing 
stable tracks of the stau in the muon chambers. We study the event 
characteristics of the long-lived staus at the LHC and demonstrate that the 
stau tracks can be distinguished from the muonic ones through proper kinematic 
cuts which also enable one to remove all standard model backgrounds.
 % end of PACS codes
} %end of abstract
\maketitle
\section{Introduction}
\label{intro}
Supersymmetry (SUSY) is still the most sought-after new physics option
around the TeV scale. 
Apart from stabilizing the electroweak symmetry breaking sector and
providing a rather tantalizing hint of Grand Unification, SUSY (in the
R-parity conserving version) also provides a cold dark matter candidate
in the form of the stable lightest supersymmetric particle (LSP). The
most common practice is to assume that the lightest neutralino is the
LSP, and the `standard' signals such as
(jets + ${E_T}\miss$), (dileptons + ${E_T}\miss$) etc. are widely studied 
under this assumption. 
However one must also know (a) how tractable are SUSY
signals with other types of LSP? And (b) is it possible to have
quasi-stable but cosmologically allowed charged particles in a SUSY
scenario, with distinct collider signatures? 

The shortest step beyond the standard electroweak
model, which provides an explanation of neutrino mass, consists in 
hypothesizing a right-handed neutrino for each fermion family \cite{rneut}.
At the same time depending on SUSY for the stabilization
of the electroweak scale,  three additional members are added to
the assortment of R-odd (super)particles, namely, the right-chiral
sneutrinos. The non-interactive nature of the dominantly right-chiral state 
(${\snu}_1$) relative to the left-chiral one makes it possible for the former 
to be a dark matter candidate \cite{arkani}. 
We study a SUSY spectrum with such a sneutrino as the LSP. 
For the the next-to-lightest SUSY particle (NLSP) one can have  
a slepton NLSP (especially the stau, with its opportunity to have a rather 
light mass eigenstate with large $\tan \beta$, $\tan\beta$ being the ratio of 
the vacuum expectation values of the two Higgs doublets), out of the many 
possibilities. 
Interestingly, for a ${\snu}_1$ LSP, any NLSP will decay into it
only through an interaction strength proportional to the neutrino mass.
This is because Yukawa interactions will invariably involve this mass,
while gauge interaction will depend on the admixture of the $SU(2)_L$
doublet component in ${\snu}_1$ proportional to the neutrino
Yukawa coupling. Thus the decay of the NLSP to the LSP is too
suppressed over most of the viable parameter space to take place within
the detector and the NLSP appears to be stable, as far as collider
detectors are concerned. We emphasize the following in this work: 
(a) The stau can become an NLSP, and the corresponding right
sneutrino, the LSP, in rather natural regions of the SUGRA parameter
space, provided that right sneutrino mass is allowed to evolve from the
common scalar mass at high scale.
(b) The stau NLSP is likely to leave charged tracks on reaching the
muon detector.

We wish to point out that the very kinematic properties of the tracks in the 
muon chamber set the long-lived staus apart from the muons in a conspicuous 
fashion \cite{gupta}, and the characteristic signal events of such a scenario 
can be separated from the standard model (SM) backgrounds in a rather 
straightforward manner.

\section{Right sneutrino LSP in supergravity}
\label{sec:1} 
The simplest extension to the SM spectrum to give Dirac masses to neutrinos,
by adding right-handed neutrino for each generation would imply
that the neutrino Yukawa couplings are quite small ($\sim 10^{-13}$).
The superpotential of the minimal SUSY standard model (MSSM)
is extended by just one term which, for a particular family,
is of the form
\bea
W_\nu^R=y_\nu \hat{H}_u \hat{L}\hat{\nu}^c_R
\eea
where $y_\nu$ is the Yukawa coupling, $\hat{L}$ is the left-handed lepton
superfield and $\hat{H}_u$ is the Higgs superfield responsible for giving
mass to the $T_3=+1/2$ fermions. The above term in the superpotential
obviously implies the inclusion of right-handed sneutrinos in the particle 
spectrum which will have all their
interactions proportional to the corresponding neutrino masses. So
the dominantly right-handed eigenstate of the tau-sneutrino might become a
possible candidate for the LSP in the
framework of minimal supergravity (mSUGRA) model of SUSY, consistent with all 
experimental bounds \cite{expbound} and also within the acceptable limits of 
dark matter density in the universe \cite{wmap,Moroi2}.
The decay rate of the stau-NLSP is extremely suppressed because of the 
smallness of the Yukawa coupling and plays a crucial 
role in our understanding of the spectrum and its consequent features.

Upon inclusion of right-chiral neutrino superfield into the SUGRA fold,
the superparticle spectrum mimics the mSUGRA spectrum in all details except
for the identity of the LSP. The mass terms for sneutrinos (neglecting 
inter-family mixing) are given by
\bea
-{\mathcal L}_{soft} \sim {M}_{{\rsnu}}^2
|\widetilde{\nu}_R|^2 + (y_{\nu}{A}_\nu H_u.\widetilde{L}
\widetilde{\nu}_R^c ~~ + ~~h.c.)
\eea
where $A_\nu$ is the term driving left-right mixing in the
scalar mass matrix, and is obtained by running of the
trilinear soft SUSY breaking term $A$. One expects minimal left-right mixing 
of sneutrinos as the Yukawa couplings are all extremely small.
The mass-squared matrix for the sneutrino thus looks like
\bea
m_{\tilde{\nu}}^2 = \left ( \begin{array}{cc} {M}_{\tilde{L}}^2+
\frac{1}{2}m_Z^2\cos2\beta & y_\nu v({A}_\nu\sin\beta-\mu\cos\beta)\\
y_\nu v({A}_\nu \sin\beta-\mu \cos\beta) &
{M}_{\tilde{\nu}_R}^2 \end{array} \right) \nonumber
\eea
where ${M}_{\tilde{L}}$ is the soft scalar mass for the
left-handed sleptons whereas the ${M}_{\tilde{\nu}_R}$ is
that for the right-handed sneutrino. In general,
 ${M}_{\tilde{L}} \ne {M}_{\tilde{\nu}_R}$ because of their different
evolution patterns as well as the D-term contribution for the former.
While the evolution of all parameters of minimal SUSY
remain practically unaffected in this scenario, the
right-chiral sneutrino mass parameter evolves \cite{arkani} at
the one-loop level as:
\bea
\frac{dM^2_{\rsnu}}{dt} = \frac{2}{16\pi^2}y^2_\nu~A^2_\nu ~~.
\eea
The extremely small Yukawa couplings cause  ${M}_{\tilde{\nu}_R}$
to remain nearly frozen at the value $m_0$, whereas the other sfermion
masses are jacked up at the electroweak scale. Thus, for a wide range of
values of the gaugino mass, one naturally has sneutrino LSP's, which,
for every family, is dominated by the right-chiral state:
\bea
\tilde{\nu}_1 = - \tilde{\nu}_L \sin\theta + \tilde{\nu}_R \cos\theta
\eea
The mixing angle is clearly suppressed due to the smallness of $y_\nu$. 
However, of the three charged slepton families, the amount of left-right
mixing is always the largest in the third (being, of course,
more pronounced for large $\tan\beta$), and the lighter stau
(${\stau}_1$) often turns out to be the NLSP in such a scenario.
Thus the mSUGRA parameter set ($m_0,m_{1/2},A,sign(\mu)~{\rm and}~\tan\beta$)
in an R-parity conserving scenario can eminently lead to a spectrum where
the right-sneutrinos will be either stable or metastable but very long-lived 
and gives a SUSY spectrum with a $\stau$ NLSP.
\begin{table}[htb]
\begin{center}
\begin{tabular}{|c|c|c|}
\hline\noalign{\smallskip}
{\bf Input}& {\bf BP-1}&{\bf BP-2} \\
\noalign{\smallskip}\hline\noalign{\smallskip}
            &$m_0=100~GeV,$
            &$m_0=110~GeV,$\\
mSUGRA      &$m_{1/2}=600~GeV$
            &$m_{1/2}=700~GeV$\\
            &$\tan\beta=30$ &$\tan\beta=10$\\
\noalign{\smallskip}\hline\noalign{\smallskip}
$\mel,\mml$                          &420&486\\%\hline
$\mer,\mmr$                          &251&289\\%\hline
$m_{\snu_{eL}},m_{\snu_{\mu L}}$     &412&479\\%\hline
$m_{\snu_{\tau L}}$                  &403&478\\%\hline
$m_{\snu_{iR}}$                      &100&110\\%\hline
$m_{\stau_1}$                        &187&281\\%\hline
$m_{\stau_2}$                        &422&486\\
\noalign{\smallskip}\hline
$m_{\chi^0_1}$                       &243&285\\%\hline
$m_{\chi^0_2}$                       &469&551\\%\hline
$m_{\chi^{\pm}_1}$                   &470&552\\%\hline
$m_{\tilde{g}}$                      &1366&1574\\
$m_{\tilde{t}_1}$                    &984&1137\\%\hline
$m_{\tilde{t}_2}$                    &1176&1365\\%\hline
$m_{h^0}$                            & 118&118\\%\hline
\noalign{\smallskip}\hline
\end{tabular}
\caption {\small \it Proposed benchmark points for study of stau-NLSP scenario
in the SUGRA fold with right-sneutrino LSP. We choose $A=100~GeV$ and 
$sgn(\mu)=+$ for both benchmark points.}
\label{tab:1}       % Give a unique label
\end{center}
\end{table}
%=============================================================
Using the spectrum generator of the package ISAJET 7.69 \cite{isajet}, we 
find that a large mSUGRA parameter space can realize this scenario of a 
right-sneutrino LSP and stau NLSP, provided that $m_0 < m_{1/2}$ and one has
$tan\beta$ of the order of 10 and above, the latter condition being
responsible for a larger left-right off-diagonal term in the
stau mass matrix (and thus one smaller eigenvalue). In Table~\ref{tab:1} we 
present two benchmark points, Benchmark Point 1 (BP-1) and Benchmark Point 2 
(BP-2) for our study of such long-lived staus at the LHC.

\section{Signatures of stau-NLSP at LHC}
In this section we discuss the signatures of the long-lived
stau-NLSP at the LHC and concentrate on two different final states,
viz.
\begin{itemize}
\item $2 \stau_1 + 2({\rm or~more})~jets~(p_T > 100~{\rm GeV})$
\item $2 \stau_1 + {\rm dimuon} + 2({\rm or~more})~jets~
(p_T > 100~{\rm GeV})$
\end{itemize}
Keeping the above signals in mind, we focus on the two benchmark points
listed in Table~\ref{tab:1} and study their signatures at the LHC,
for an integrated luminosity of $30~fb^{-1}$.

The $\stau_1$-NLSP is long-lived and stable in the
context of collider studies. So it will almost always decay outside
the detector, leaving characteristic signals like charged tracks,
with large transverse momenta. In fact this would be quite a contrast to the
traditionally thought of SUSY signals with {\it large missing 
transverse energy} and the stable stau
will behave just like a muon. However, in the absence of spin identification,
these staus will behave more like very heavy muon-like particles
with $\beta(=v/c)< 1$ and such heavy charged particles will
%========================================================================
\begin{figure*}[htb]
\begin{center}
\includegraphics[width=0.23\textwidth,height=0.23\textwidth,angle=0]{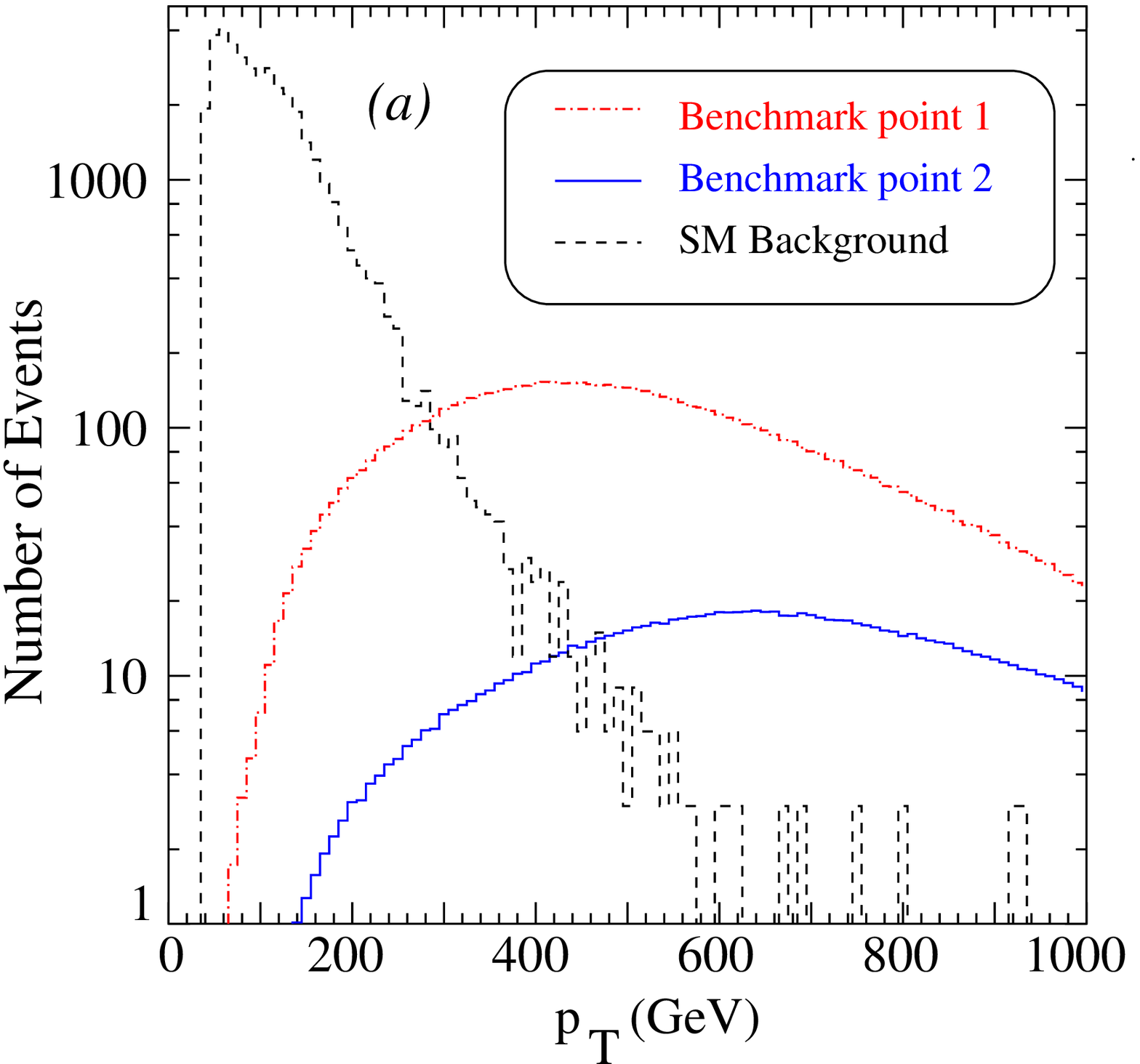}
\includegraphics[width=0.23\textwidth,height=0.23\textwidth,angle=0]{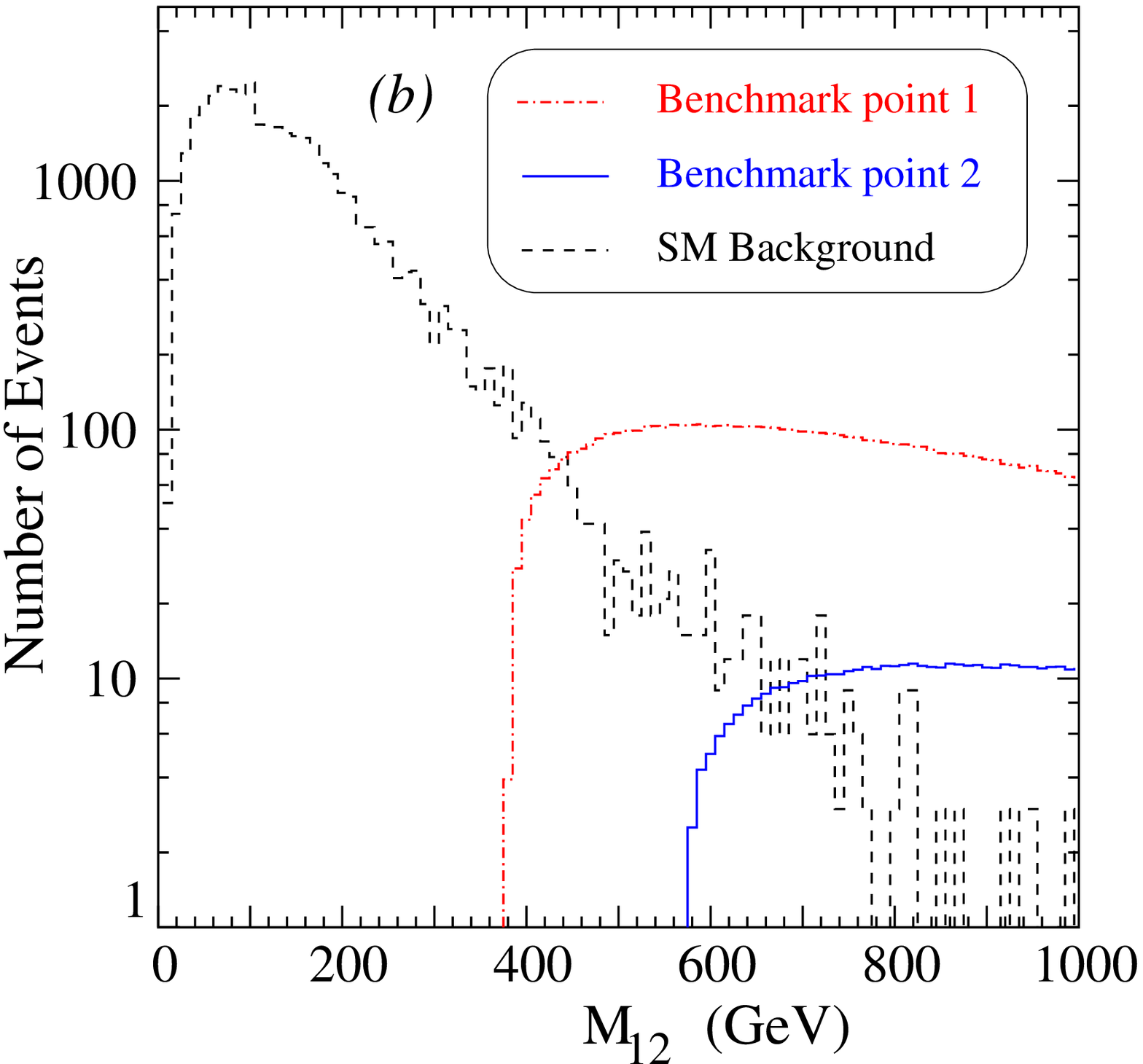}
\includegraphics[width=0.23\textwidth,height=0.23\textwidth,angle=0]{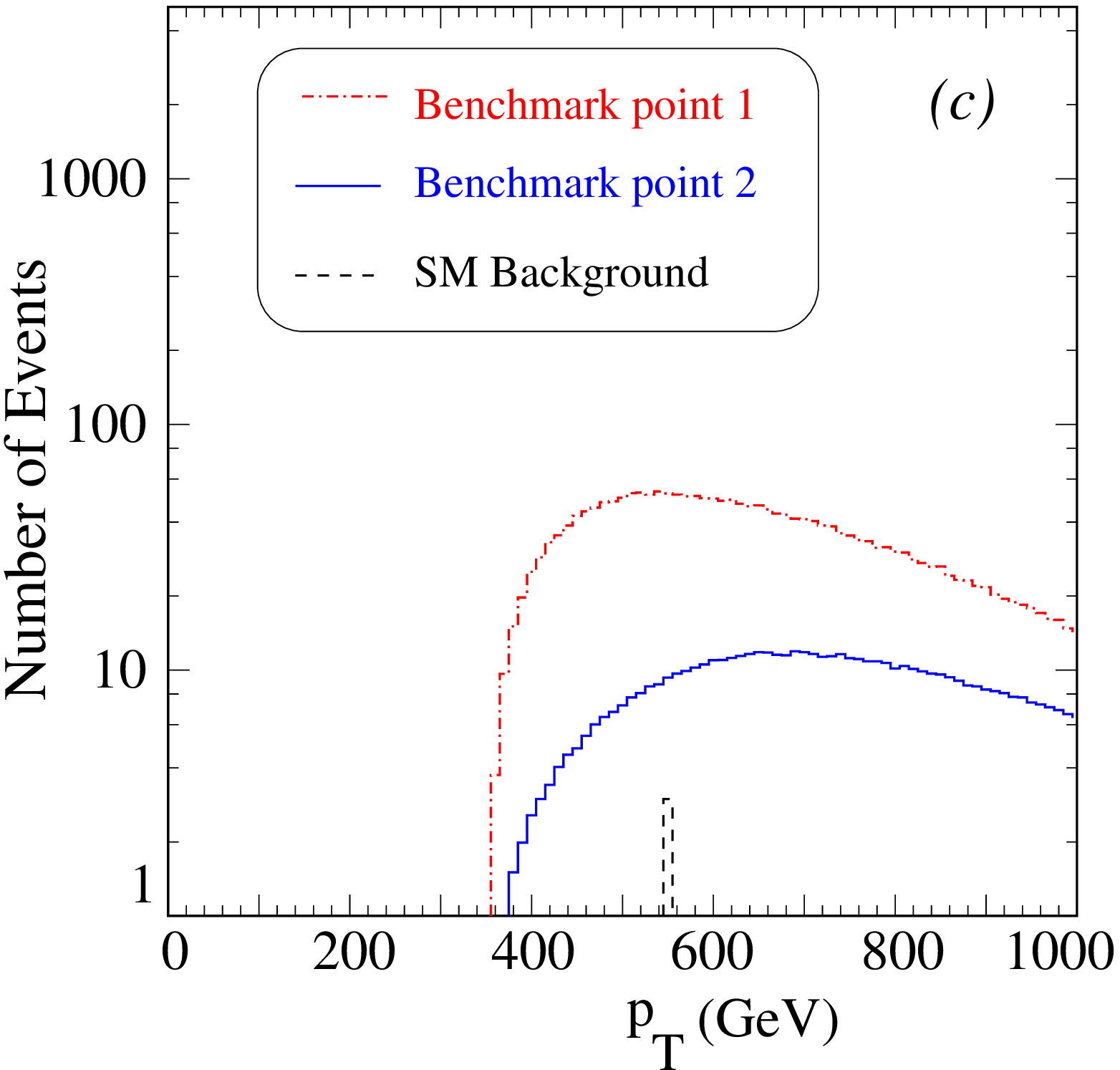}
\includegraphics[width=0.23\textwidth,height=0.23\textwidth,angle=0]{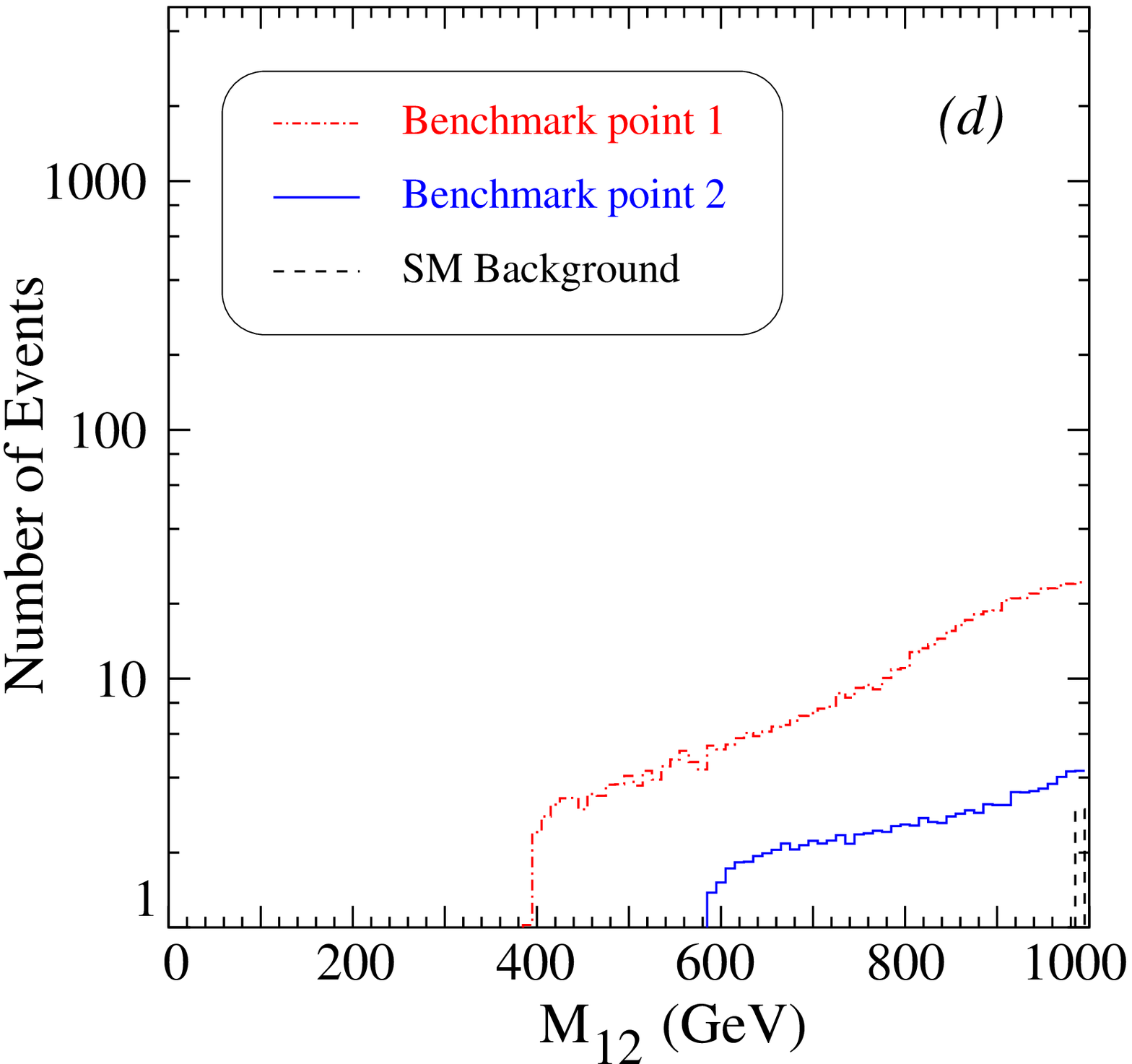}
\end{center}
\caption{\small
\it{Kinematic distributions for the signal
$2\stau_1 + (\ge 2)$ hard jets. In (a) the transverse momentum
distributions for the harder $\stau_1$ is shown and (b) shows the
invariant mass distribution for the $\stau_1$ pair. 
(c) and (d) show the same distributions after imposing the stronger cuts.}}
\label{fig:1}       % Give a unique label
\end{figure*}
%========================================================================
have high specific ionisation due to their slow motion within the
detector. In this work we take a qualitatively different approach, 
which would be based more on an analysis
pertaining to studying the kinematics of processes producing such
particles at the LHC. \\
$\bullet$ {\bf {$2 \stau_1$ + hard jets}} \\
The signals mentioned here arise mostly from
the direct decay of gluinos and squarks, produced via strong interaction at
the LHC, into the lightest neutralino, with the latter decaying into a tau 
and a lighter stau, and the tau decaying hadronically in turn. 
Cascades through other neutralinos and charginos supplement the rates to a 
moderate extent. We have used PYTHIA 6.409 \cite{pythia} for our event 
generation and interfaced it with ISASUGRA, contained in ISAJET 7.69, 
to generate the mSUGRA spectrum with a right-sneutrino LSP with its 
corresponding renormalization group equation RGE (eqn. 3) also included.

The parton densities have been evaluated at $Q=2m_{\tilde{\tau}_1}$
using CTEQ5L \cite{cteq},
and the renormalization scale and factorization scale are
$\mu_F=Q=\mu_R$.
The effects of both initial state radiation (ISR) and final
state radiation (FSR) along with the effects of hadronization and multiple 
%------------------------------------------------------------------
\begin{table}[htb]
\begin{center}
\begin{tabular}{|c|c|c|c|}
\hline\noalign{\smallskip}
{\bf{Cuts}}& {\bf{SM}}&{\bf{BP-1}}&{\bf{BP-2}}\\
\noalign{\smallskip}\hline\noalign{\smallskip}
Basic& 39617&8337&1278\\\hline
Basic + & &&\\
$p_T>350$ GeV& 5&2587&737\\
\noalign{\smallskip}\hline
\end{tabular}
\caption{\small
\it{The expected number of events for the signal
and background with the different cuts imposed on the selection of
events.}}
\label{tab:2}
\end{center}
\end{table}
%-----------------------------------------------------------------------
interaction with the help of PYTHIA hadronization schemes are included.
To define jets we use the simple-minded jet cone algorithm implemented
in PYTHIA through the subroutine PYCELL. 
To select our final states, we demand the following requirements
(called basic cuts) on our sample events:
\begin{itemize}
\item Each $\stau_1$ should have $p_T > 30$ GeV and at least two jets 
with $p_T > 100$ GeV (hard jets).
\item Both the $\stau_1$'s should satisfy $|\eta|\leq 2.5$, to ensure
that they lie within the coverage of the muon detector.
\item  $\Delta R_{\stau_1\stau_1} \geq 0.2$, to ensure that the $\stau_1$'s
are well resolved in space.
\item In addition we have rejected events having photons with
$|\eta_{\gamma}|\leq 2.5$ and ${p_T}_{\gamma}\ge 25$ GeV.
\end{itemize}
%========================================================================
As the charged tracks pass through the muon chamber, it is probable
that the muonic events will fake our signal. Therefore, events with
two or more hard jets and two central muons will {\it prima facie}
constitute our standard model background. The leading contribution
to such final states satisfying our basic cuts comes
from top-pair production and its subsequent decay into
dimuons, with similar topology as that of the signal. The
sub-leading contributions consist in weak boson pair production.
For performing the background analysis we again used the same criteria as
stated above for the signal, with all kinematic features of
the long-lived charged tracks attributed to the muons.

An important point to note is that the background is almost
completely reducible with the imposition of stronger event selection
criteria, as we shall see below.
In Figure~\ref{fig:1},
we present distributions of a few observables where we could distinguish the
signal from backgrounds. These are the transverse momentum ($p_T$) of the
harder charged track (and the harder muon in the case of backgrounds)
and the invariant mass ($M_{12}$) of the two charged tracks (or dimuons).
In addition, the radii of curvatures of the stau tracks will also lie
in a clearly distinguishable range, as seen from the $p_T$-distributions.
On examining the  $p_T$-distributions,
we find it most convenient to eliminate the background by imposing a stronger
$p_T$ cut on {\em both the charged tracks} of 350 GeV. 
This makes the signal stand out clearly for both the benchmark points, as can 
be seen from Table~\ref{tab:2} and also Figure \ref{fig:1}(c) 
and \ref{fig:1}(d). These constitute the real `signal distributions', and can 
be used to extract information, for example, about the mass of the stau-NLSP 
and other SUSY parameters.

The signals in the scenario under investigation here will be more
severely affected when a missing-$p_T$ cut is imposed.
To take a specific example, the signal rate at BP-1
becomes about 48\% of its original value when the requirement of
a minimum missing-$p_T$ of 100 GeV is imposed. In contrast, at a
a nearby point in the parameter space
($m_0~=~200$ GeV, $m_{1/2}~=~600$ GeV, $A~=~100$ GeV, $\tan\beta~=~30$,
$sgn(\mu)$ = +ve), the same missing-$p_T$ cut allows about 97\%
survival of the $jets~+~dimuons~+\not{p_T}$ signal. Thus the response
to missing-$p_T$ cuts turns out to be an effective tool of differentiation
between our signal and that coming from MSSM (with neutralino LSP), at least 
when R-parity is conserved.\\
$\bullet$ {\bf{Dimuon and 2$\stau$ and $+(\ge 2)$ jets:}}\\
With the stringent demand on the hardness of the jets, this is a very
clean signal, albeit less copious than the previous one 
(BP-1=689 events, BP-2=103 events, SM=83 events). 
Such final states will require cascade decays of gluinos and squarks involving
the charginos and heavier neutralinos. The same
`basic cuts' are imposed here, too, which are found to be sufficient
in drastically reducing the SM backgrounds in the form of four muons
together with two or more jets with $p_T > 100$ GeV (SM=29 events).
%========================================================================
\begin{figure}[htb]
\begin{center}
\includegraphics[width=0.23\textwidth,height=0.23\textwidth,angle=0]{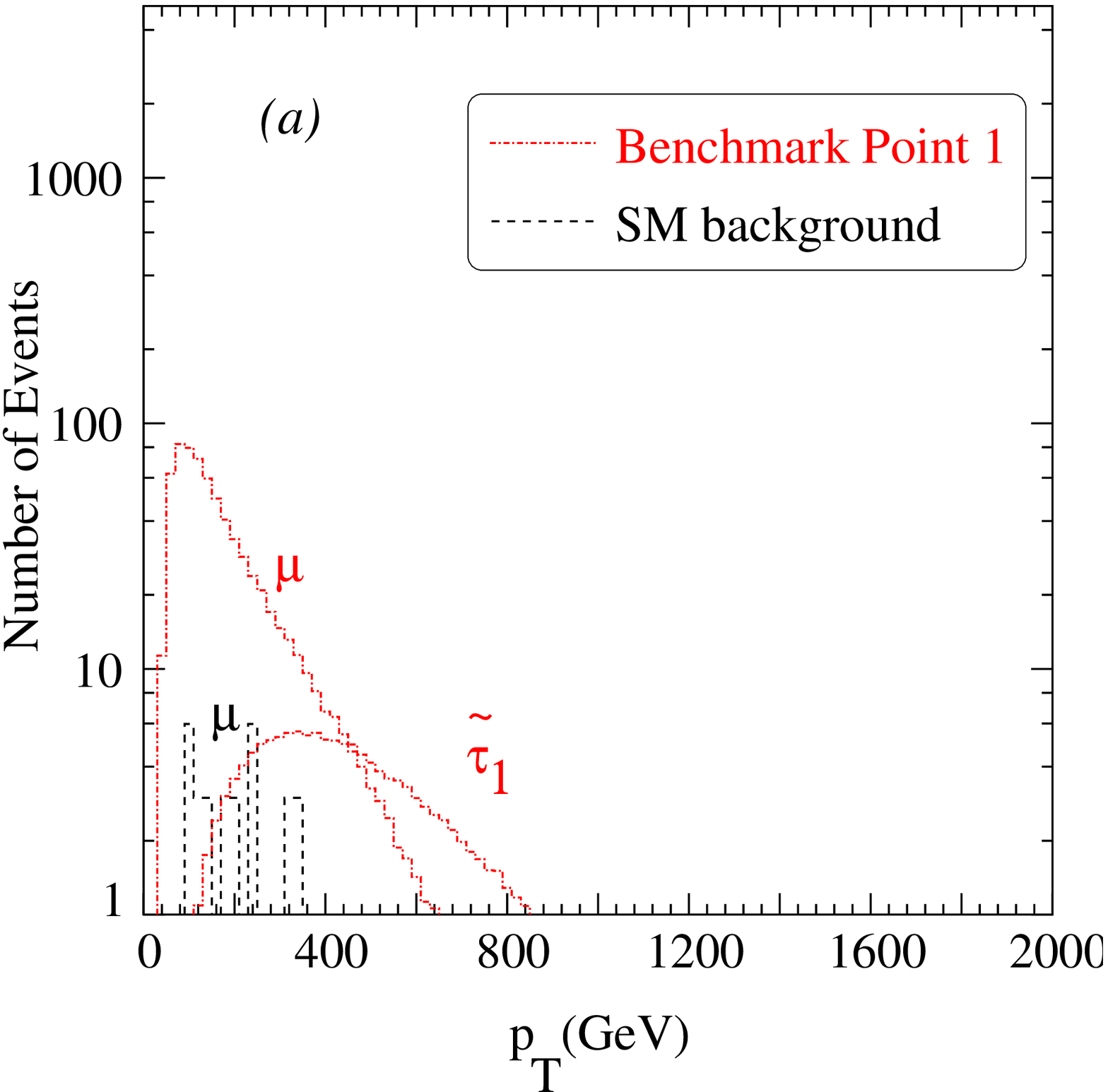}
\includegraphics[width=0.23\textwidth,height=0.23\textwidth,angle=0]{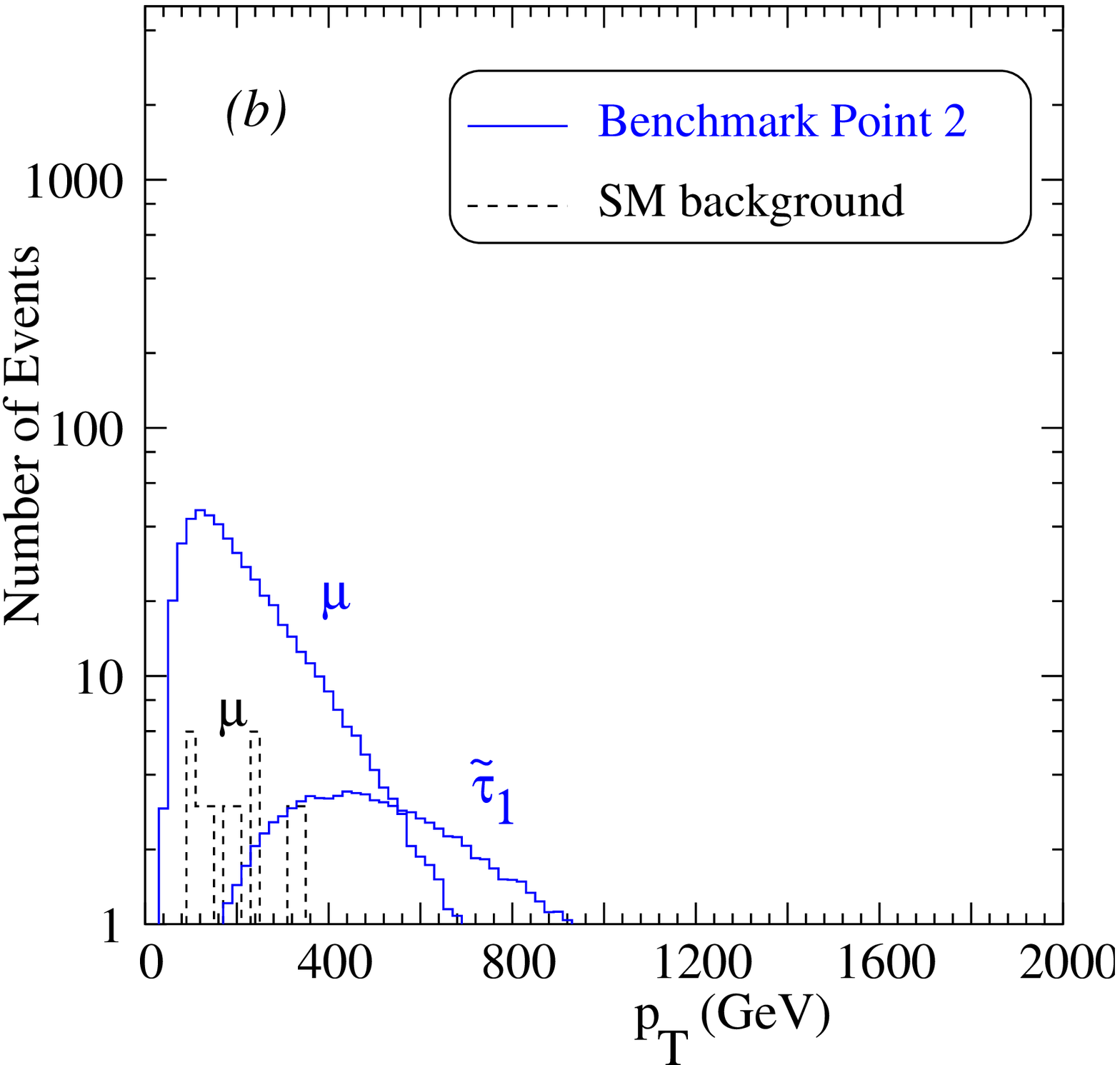}
\includegraphics[width=0.25\textwidth,height=0.23\textwidth,angle=0]{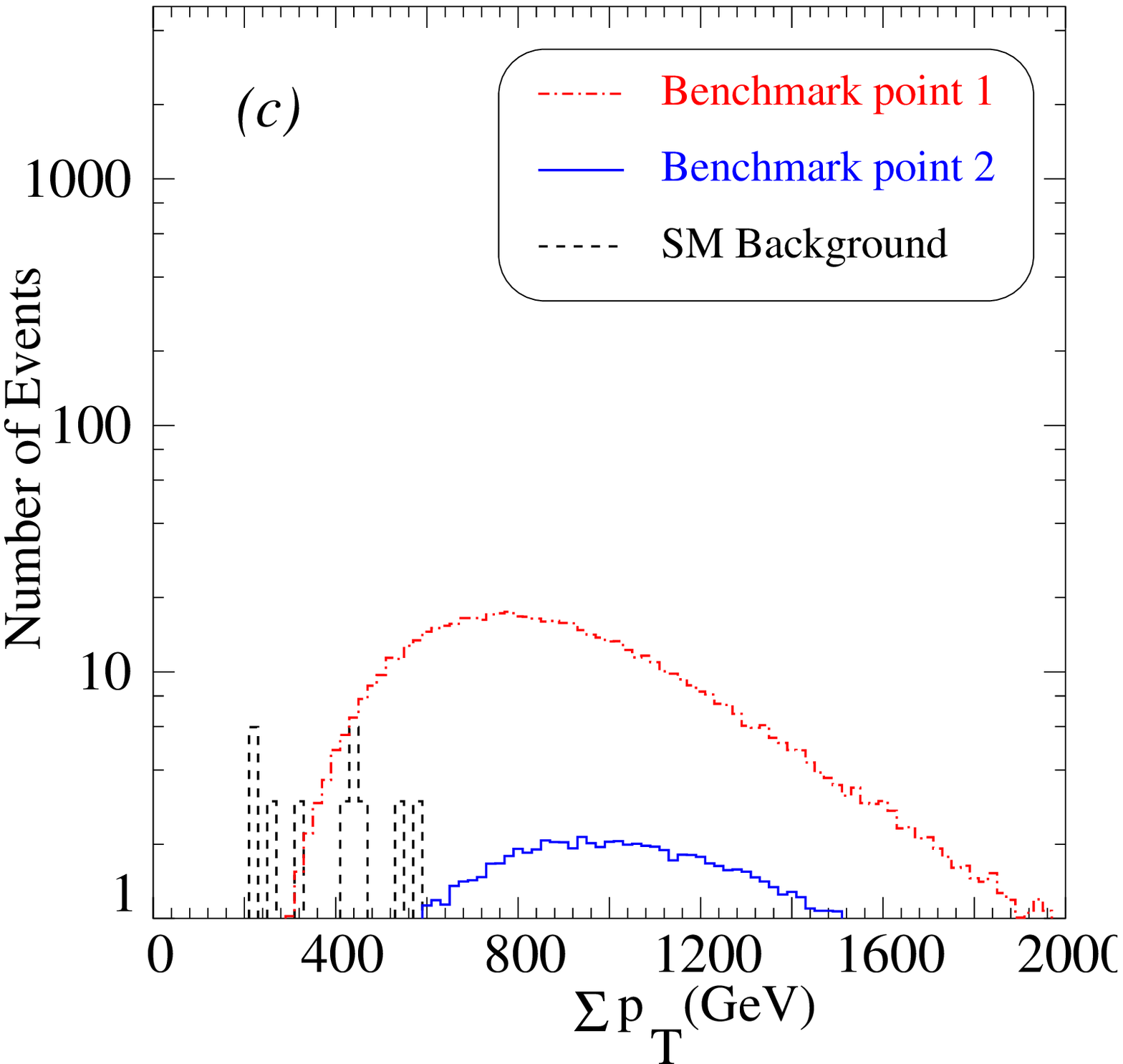}
\end{center}
\caption{\small
\it{Kinematic distributions for the signal
$2\stau_1 + dimuon + (\ge 2)$ hard jets with basic cuts.
In (a) and (b) the transverse momentum
distributions for the harder $\stau_1$ and harder muon of the signal
is shown and (c) shows the scalar sum of $p_T$'s of (dimuon +
2$\stau_1$) of the signal. We follow the same notation as in Figure
\ref{fig:1}.}} 
\label{fig:2}       % Give a unique label
\end{figure}
%========================================================================
In Figure~\ref{fig:2} we plot the corresponding distributions for the above
final states. Furthermore, the plot of the scalar sum of the individual 
transverse momenta of the charged tracks in the muon chamber, in
Figure~\ref{fig:2}(c), indicates that
a cut of 600 GeV on this sum washes out the backgrounds completely leaving
553 events for BP-1 and 89 events for BP-2.
The advantage of applying this cut is that there is no loss
of signal events in the case of a relatively heavy NLSP.
In Figure~\ref{fig:2}(a) and (b), the $p_T$-distributions of the
(harder) muon and the corresponding stau-track are seen to have a
substantial overlap. Therefore, a distinction between them based on
the thickness of the tracks as well as the information provided
by measurement of the `time-of-flight' can be useful here.
%-----------------------------------------------------------------------
\section {\bf{Summary and conclusions}}
We have studied a SUSY scenario with a stau-NLSP and an
overwhelmingly right-chiral sneutrino as the LSP, where
the sneutrino is at least partially responsible for the
cold dark matter of the universe. A mass spectrum corresponding
to such a scenario can be motivated in a SUGRA framework,
including right-chiral sneutrinos whose masses remain practically
frozen at the universal scalar mass at the SUSY breaking scale.
We find that the superparticle
cascades culminating into the production of stau-pairs give
rise to very distinct signals of such a scenario. Although the
charged tracks of the quasi-stable staus tend to fake
muonic signals in the muon chamber, our analysis reveals
considerable difference in their kinematic characters.
Such difference can be used in a straightforward way
to distinguish between the long-lived staus and the muons,
and also to eliminate all standard model backgrounds. Since
the mass spectrum under consideration here is as probable
as one with a neutralino LSP in mSUGRA, further study of
all possible ways of uncovering its signature at the LHC
should be of paramount importance.

This talk was based on the work done in Ref~\cite{gupta}.


\begin{thebibliography}{999}
\bibitem{rneut}
  A.~T.~Alan and S.~Sultansoy,
  %``The 'right' sneutrino as the LSP,''
  J.\ Phys.\ G \textbf{30}, (2004) 937; 
  T.~Asaka, K.~Ishiwata and T.~Moroi,
  %``Right-handed sneutrino as cold dark matter,''
  Phys.\ Rev.\ D \textbf{73}, (2006) 051301. 
\bibitem{arkani}
  N.~Arkani-Hamed, L.~J.~Hall, H.~Murayama, D.~R.~Smith and N.~Weiner,
  %``Small neutrino masses from supersymmetry breaking,''
  Phys.\ Rev.\ D \textbf{64}, (2001) 115011;
  D.~Hooper, J.~March-Russell and S.~M.~West,
  %``Asymmetric sneutrino dark matter and the Omega(b)/Omega(DM) puzzle,''
  Phys.\ Lett.\ B \textbf{605}, (2005) 228.
%\cite{Gupta:2007ui}
\bibitem{gupta}
  S.~K.~Gupta, B.~Mukhopadhyaya and S.~K.~Rai,
  %``Right-chiral sneutrinos and long-lived staus: Event characteristics at the
  %Large Hadron Collider,''
  Phys.\ Rev.\  D {\bf 75} (2007) 075007.
  %%CITATION = PHRVA,D75,075007;%%
\bibitem{expbound}
  W.~M.~Yao {\it et al.},
  %``Review of particle physics,''
  J.\ Phys.\ G \textbf{33}, (2006) 1;
  E.~Barberio {\it et al.}  [Heavy Flavor Averaging Group (HFAG)],
  %``Averages of b-hadron properties at the end of 2005,''
  arXiv:hep-ex/0603003.
\bibitem{wmap}
  D.~N.~Spergel {\it et al.},
  %``Wilkinson Microwave Anisotropy Probe (WMAP) three year results:
  %Implications for cosmology,''
  Astrophys.\ J.\ Suppl.\  \textbf{170} (2007) 377.
\bibitem{Moroi2}
  T.~Asaka, K.~Ishiwata and T.~Moroi,
  %``Right-handed sneutrino as cold dark matter of the universe,''
 Phys.\ Rev.\  D \textbf {75} (2007) 065001.
  %%CITATION = PHRVA,D75,065001;%%
\bibitem{isajet}
  F.~E.~Paige, S.~D.~Protopopescu, H.~Baer and X.~Tata,
  %``ISAJET 7.69: A Monte Carlo event generator for p p, anti-p p, and e+ e-
  %reactions,''
  arXiv:hep-ph/0312045.
\bibitem{pythia}
  T.~Sjostrand, S.~Mrenna and P.~Skands,
  %``PYTHIA 6.4 physics and manual,''
  JHEP \textbf{0605}, (2006) 026.
\bibitem{cteq}
  H.~L.~Lai {\it et al.},
  %``Global QCD Analysis And The Cteq Parton Distributions,''
  Phys.\ Rev.\ D \textbf{51}, (1995) 4763.
\end{thebibliography}
\end{document}